# Integrated magnonic chip using cascaded logic

*Mengying Guo[1*], Xudong Jing[1*], Kristýna Davídková[2*], Roman Verba[3], Zhenyu Zhou[1], Xueyu Guo[1], Carsten Dubs[4], Yiheng Rao[5], Kaiming Cai[1], Jing Li[6], Philipp Pirro[7], Andrii V. Chumak[2], Qi Wang[1†]*

[1] School of Physics, Hubei Key Laboratory of Gravitation and Quantum Physics, Institute for Quantum Science and Engineering, Huazhong University of Science and Technology, Wuhan, China

[2] Faculty of Physics, University of Vienna, Vienna, Austria

[3] V. G. Baryakhtar Institute of Magnetism of the NAS of Ukraine, Kyiv, Ukraine

[4] INNOVENT e.V., Technologieentwicklung, Jena, Germany

[5] School of Microelectronics, Hubei University, Wuhan, China

[6] Wuhan National High Magnetic Field Center, Huazhong University of Science and Technology, Wuhan, China

[7] Fachbereich Physik and Landesforschungszentrum OPTIMAS, Rheinland-Pfälzische Technische Universität Kaiserslautern-Landau, Kaiserslautern, Germany

The transistor transformed not only electronics but everyday life, and the integrated circuit - now simply the "chip" - made computation scalable and ubiquitous. Magnonics has long promised a parallel path to low-energy information processing by using spin waves instead of charge. Progress, however, has been limited by two fundamental obstacles: intrinsic attenuation of spin waves and the requirement for precisely normalised output intensity and input phase to ensure reliable logic operation - conditions that are difficult to maintain in large-scale circuits owing to inevitable imperfections. Here, we report an integrated magnonic circuit that overcomes both limitations through engineered nonlinearity in nanoscale yttrium iron garnet waveguides. Nonlinear self-adjustment of the spin wave phase renders logic operation insensitive to the relative phases of the inputs, while a deeply nonlinear, threshold-activated self-normalised excitation restores and standardises the output intensity. Using space-resolved micro-focused Brillouin light scattering, we demonstrate reconfigurable AND, OR and three-input majority gates and realise deterministic cascading across sequential stages, establishing a scalable on-chip logic primitive. The architecture operates with gigahertz frequencies, supports dynamic threshold control for functional reconfiguration, and is compatible with scalable integration, making it attractive for adaptive and neuromorphic computing. By resolving phase-independent operation and signal restoration at the level of device physics, this work advances magnonics from isolated proof-of-concept devices towards integrated magnonic chips that can complement advanced CMOS in energy-constrained computing tasks.

---

[*] These authors contributed equally to this work.
[†] Corresponding Author: williamqiwang@hust.edu.cn

s

**Introduction**

The integrated circuit transformed transistor physics into a scalable platform that reshaped industry and daily life. By integrating many active elements on a single chip, it enabled compact, affordable, and reliable computation at a scale that unattainable with discrete devices. An analogous inflection point is now within reach for spin-wave-based information processing. Magnonics replaces moving charge with collective spin excitations and promises low-energy operation, native compatibility with microwave signals, and access to rich nonlinear phenomena [1-5]. Despite these advantages, the field has so far lacked an analogue of the integrated circuit. Although individual magnonic devices have demonstrated impressive functionality [6-12], including recent AI-driven inverse design devices for logic [13,14] and neuromorphic computing [15-17], these implementations remain largely isolated rather than forming scalable, integrated architectures.

A major challenge is that many magnonic logic schemes rely on phase encoding, which makes their operation sensitive to phase variations [18-20] and hinders reliable cascading, thereby limiting practical on-chip logic [20-24]. In parallel, magnetic damping causes substantial signal attenuation during spin-wave propagation, so that downstream gates require active signal restoration through normalization rather than simple amplification [8,23,25-27]. An integrated magnonic chip that simultaneously resolves phase tolerance and intrinsic signal normalization would therefore represent a decisive step from isolated proof-of-concept demonstrations toward a viable technology, and opening a credible path to energy-efficient computing that complements advanced CMOS in wave-based tasks.

Here we present a compact nonlinear magnonic chip based on nanoscale yttrium iron garnet (YIG) that exploits engineered spin-wave nonlinearity to achieve phase-insensitive logic operation and intrinsically normalised output intensities, making the platform suitable for scalable integration. Reconfigurable AND and OR logic operations are demonstrated in a Y-shaped magnonic gate, and the same physical architecture is extended to realise a three-input majority gate. The underlying nonlinear dynamics are clarified through a detailed analysis of the phase and intensity evolution of spin waves within the Y-shaped structure. Reliable gate-to-gate operation is established by demonstrating deterministic state transfer in two cascaded logic gates. The architecture is compatible with wafer-scale fabrication and heterogeneous integration, supporting dense magnonic logic and programmable neuromorphic computing with microwave frequencies.

**Results**

Figure 1a presents a scanning electron microscopy (SEM) image of a Y-shaped gate, with 800-nm-wide waveguides fabricated from a 103-nm-thick YIG film [28,29] using electron-beam lithography and hard-mask etching (see Methods). Three gold stripe antennas are patterned on top: two as inputs and one as a pump. As indicated in Fig. 1a, the antenna widths are 3 μm and 2.5 μm, respectively. An external magnetic field of 330 mT is applied along the z-axis to saturate the magnetization out-of-plane, allowing the use of forward volume spin waves that support a strong nonlinear behaviour [27,30,31]. Microwave pulses (0.8 μs duration, 1 μs period) at 5.6 GHz are applied to both input and the pump antennas. The microwave power is adjusted so that spin waves are directly excited at the inputs (15 dBm), while the pump antenna is operated within its bistable region, where the magnon state (low or high intensity) depends on its excitation history [31-33]. At this frequency, the 2.5 μm-wide pump antenna exhibits low excitation efficiency and cannot directly generate spin waves at low pump power [34,35]. However, when spin waves from the inputs reach the pump region, they perturb the local magnonic ground state beneath the pump antenna. Once the incoming spin-wave intensity exceeds a certain threshold, the pump antenna is triggered and begins to re-emit spin waves. This nonlinear re-emission process cleans up the original signal, improving its quality and amplifying the retransmitted spin wave to a fixed output intensity [31]. This makes the output suitable for cascading into subsequent logic elements in an integrated magnonic circuit. Moreover, this trigger mechanism exhibits clear threshold behaviour, which forms the basis for realizing magnonic logic gates.

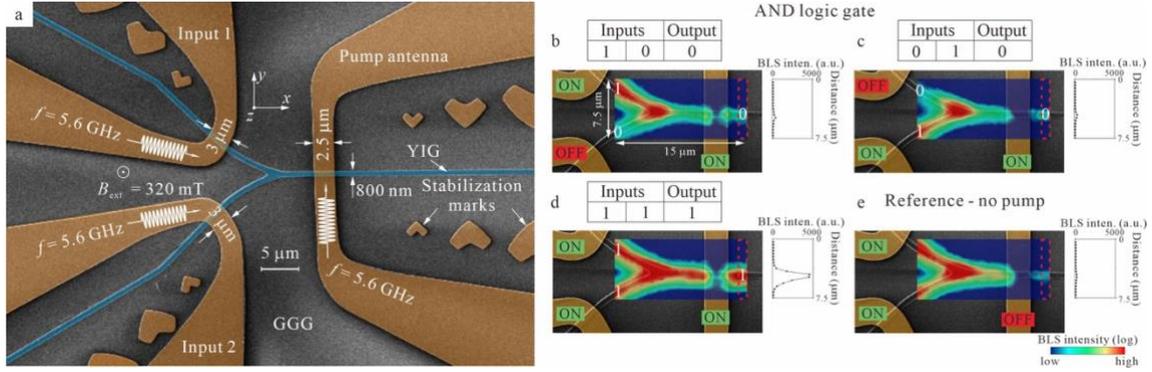

*Figure 1.* **Sample geometry and operational principle of Y-shape magnonic AND logic gate.** *a. Scanning electron microscopy (SEM) image of the Y-shaped magnonic logic gate (shaded in blue). The device features three gold stripe antennas, which function as two input ports and a pump port, as indicated. An external magnetic field of 330 mT is applied along the out-of-plane direction (z-axis) to saturate the magnonic waveguide, establishing forward volume configuration. Spin waves are generated by applying a 5.6 GHz microwave current to the two input antennas, while the pump antenna is driven with the same frequency signal to provide spin-wave intensity-coded logic operation and amplification. Panels (b-e) show two-dimensional (2D) Brillouin light scattering (BLS) maps of the integrated BLS intensity (4.8-6.0 GHz) for different logic inputs. The maps were obtained by scanning the laser spot over a 15×7.5 μm² area with a resolution of 40×20 points. The specific input configurations are: (b) '10', (c) '01', (d) '11', and (e) a reference state without a pump signal. The right panels show the spin-wave intensity integrated over the red dashed rectangular regions at the end of the output. The reduced BLS intensity above the pump antenna is due to optical shadowing by the metallic antenna, which prevents detection of spin waves directly underneath.*

Building on this nonlinear, threshold-activated response, logical information is encoded in the spin-wave intensity, with low and high intensity representing logic '0' and logic '1', respectively. To implement the AND logic operation, where the output is activated only when both inputs are logic '1', the microwave power applied to the pump antenna is set to a specific value of 4 dBm. This optimized power ensures that the operating frequency of 5.6 GHz falls within the bistable window of the pump antenna [see Supplementary Materials S1]. Under these conditions, the pump antenna is activated only when spin waves from both input antennas are present, thereby preventing excitation by a single input or by the pump alone. The functionality of the magnonic AND gate is directly demonstrated by two-dimensional (2D) Brillouin light scattering (BLS) spectroscopy [36] scans of the spin-wave intensity, shown in Fig. 1b-d. These maps display the integrated BLS intensity within the frequency range of 4.8 to 6.0 GHz for different input logic states. The input configurations ('10', '01', '11') are indicated above each panel. The panels on the right quantify the spin-wave intensity integrated within the red dashed rectangular regions near the output. These data confirm the AND-gate logic functionality: the output intensity remains low (logic '0') for the input combinations '10' and '01' (Fig. 1b, c), and becomes high (logic '1') only when both inputs are active ('11', Fig. 1d). The reduced BLS intensity observed above the pump antenna arises from optical shadowing by the metallic antenna, which prevents detection of spin waves directly beneath it. Figure 1e shows a reference measurement with both input antennas active and the pump antenna switched off, resulting in a low output signal. This confirms that the spin waves detected in the logic '1' state are not due to the direct propagation from the inputs, which undergo attenuation, but are instead predominantly generated by the pump antenna through nonlinear re-emission triggered by incoming spin waves.

Crucially, the switching threshold of the bistable pump antenna is not fixed but programmable via the pump power. This tunability enables dynamic reconfiguration of the logic functionality. For instance, when the pump power is increased to 9 dBm, the antenna remains unable to excite spin waves alone but becomes responsive to either input, thereby realizing an OR logic gate within the same device [see Supplementary Materials S1 and S2]. This ability to continuously tune the activation threshold highlights that the underlying mechanism is not limited to binary gating, but rather provides a general threshold-based nonlinear activation process that, e.g. are also utilized for neuromorphic computing [15-17].

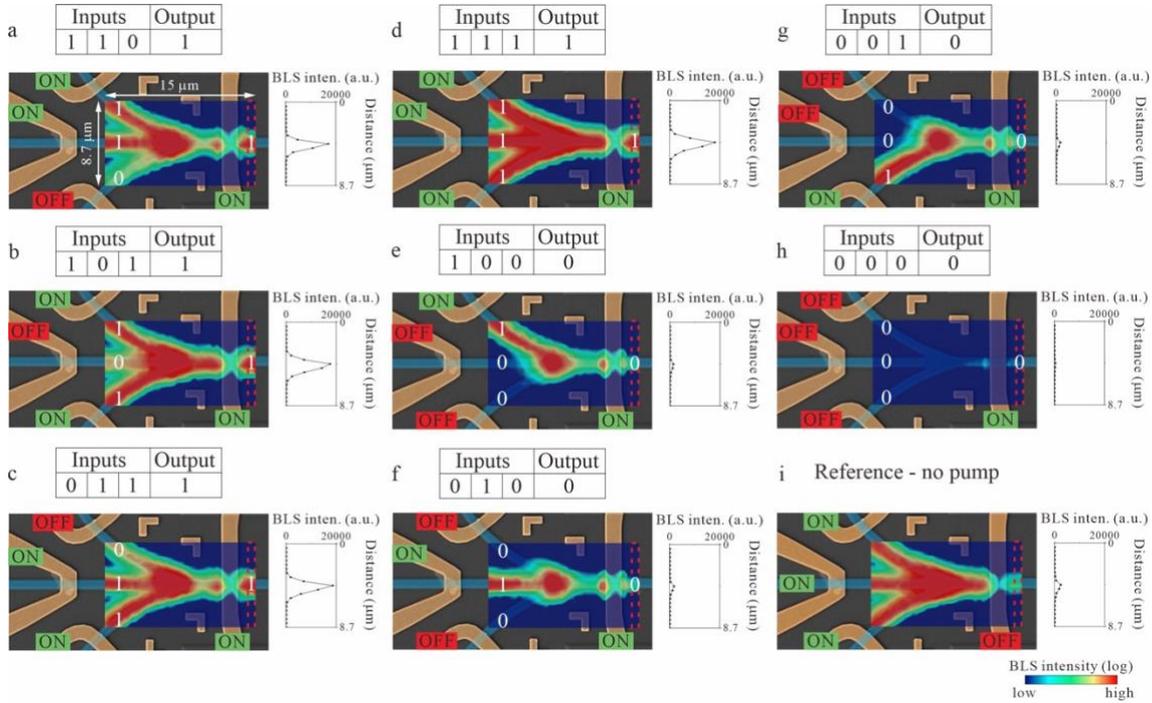

*Figure 2 **Functionality of magnonic majority gate.** Two-dimensional Brillouin light scattering intensity maps illustrating the spin-wave distribution within the three-input magnonic majority gate for different input logic states. Three input antennas are symmetrically coupled to a bistable pump antenna whose switching threshold determines the output state. When at least two inputs are in the logic '1' state, the total spin-wave intensity exceeds the activation threshold, and the pump antenna switches to the high-emission state, producing a strong output signal corresponding to logic '1' (a-d). When fewer than two inputs are active, the combined excitation remains below threshold, and the antenna stays in the low-emission state, resulting in logic '0' (e-h). The right panels display the spin-wave intensity integrated over the red dashed rectangular regions near the output, confirming the discrete transition between low and high emission states. i, a reference without pump signal.*

Exploiting this threshold-controlled activation, the device can be extended to logic operations that require the nonlinear combination of multiple inputs. By appropriately choosing the pump power, a similar physical structure is reconfigured into a three-input magnonic majority gate, as shown in Fig. 2. Majority-gate primitives enable compact logic synthesis, with the potential to reduce the number of components by almost an order of magnitude: for example, a full adder can be implemented using only three majority gates, whereas a conventional CMOS implementation typically requires about 28 field-effect transistors. Previous magnonic majority gates predominantly relied on phase encoding, making their operation highly sensitive to phase variations. In integrated circuits, however, only the input phase of the first stage can be precisely defined, the phases in subsequent stages are determined by the output phase of the preceding gate and the propagation distance. In practice, fabrication-induced imperfections and defects render the accumulated propagation phase difficult to control. Moreover, phase-based majority gates generally produce output signals whose intensity depends on the specific input combination, preventing reliable signal normalization and thereby hindering cascading [22,24,37,38].

Here, these long-standing limitations are overcome by encoding information in the spin-wave intensity rather than phase. The nonlinear, threshold-based activation of the pump antenna ensures that the output intensity is intrinsically normalized and insensitive to phase variations, enabling deterministic cascading of majority logic. The 2D BLS intensity maps visualize the spin-wave distribution within the gate for different input states. When at least two of the three inputs are logic '1', the combined spin-wave intensity at the pump antenna exceeds the activation threshold, triggering a transition to the high-emission state and yielding a strong output signal (Fig. 2a-d). When fewer than two inputs are active, the excitation remains below threshold, and the antenna stays in its low-emission state (Fig. 2e-h). Similar to Fig. 1e, the reference measurement in Fig. 2i confirms that the logic '1' output arises predominantly from the pump antenna's re-emission rather than direct propagation from the inputs.

The panels on the right show the spin-wave intensity integrated over the red dashed rectangular regions at the end of the output waveguide, providing a quantitative measure of the output level. Owing to the threshold-driven nonlinear activation, the output intensity is intrinsically normalized: once switched on, the pump antenna emits a well-defined and nearly constant intensity that is independent of the specific input combination. This binary-like, normalized emission solves a long-standing problem in magnonic majority gate design, where previous phase-coding-based implementations failed to produce consistent output intensity for all input combinations [22,24,37,38]. By tuning the pump power, the same physical structure can be dynamically reconfigured to implement AND, OR, or majority functions within a single device. This reconfigurable unit establishes a foundational building block for scalable magnonic circuits, potentially extendable to neuromorphic and multi-bit computational schemes [15-17].

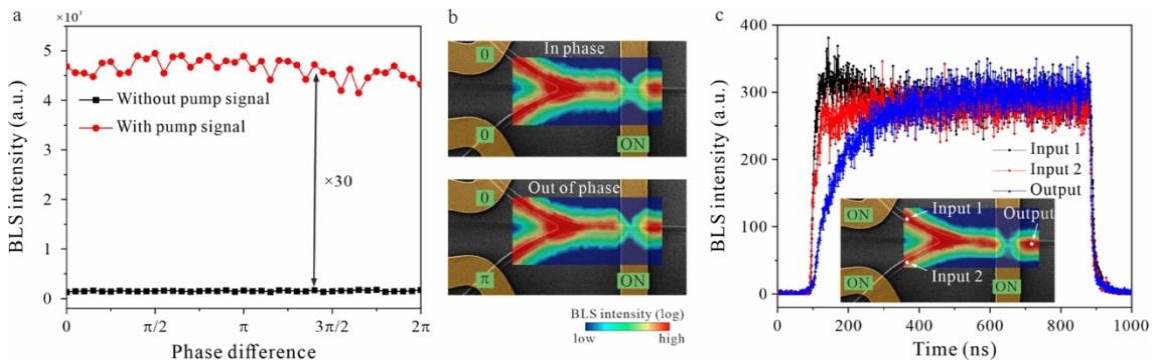

*Figure 3. **Phase-insensitive magnonic logic gate with intensity normalization.** a, Output spin-wave intensity as a function of phase difference of two inputs at the condition with (red) and without (black) pump signal. b, The two-dimensional spin-wave intensity for in phase and out-of-phase input waves. c. The spin-wave intensity as a function of time measured at two inputs (black and red) and output (blue).*

To demonstrate operation robustness and potential for the cascadability of the proposed magnonic logic gate, Fig. 3a shows the output intensity of the Y-shaped gate as a function of the input phase difference, both with (red curve) and without (black curve) the pump signal. In contrast to conventional interference-based gates, the activated device delivers an output that is insensitive to the relative input phases. This crucial property is confirmed by the 2D BLS maps in Fig. 3b, which display nearly identical spin-wave intensity distributions for in-phase and out-of-phase inputs. Such phase insensitivity is essential for cascading, as it prevents the accumulation of uncontrolled phase distortions during signal propagation across multiple logic gates. The underlying mechanism is discussed below.

An additional advantage of the proposed design is that the pump antenna performs a dual role: it operates as a threshold-activated switch and simultaneously as a normalized amplifier that compensates for propagation losses while maintaining a constant output intensity. Compared with the unpumped case, the activated pump enhances the output spin-wave intensity by approximately thirtyfold as shown in Fig. 3a. Furthermore, because the spin waves at all ports are excited via a deeply nonlinear frequency shift, the output intensity depends solely on the excitation frequency and remains unaffected by variations in input or pump power once the threshold is surpassed [27,31]. This self-normalized excitation mechanism guarantees uniform spin-wave intensities at both input and output ports - a key prerequisite for multi-stage cascading. The time-resolved measurements in Fig. 3c further confirm this behaviour: all three ports exhibit comparable signal intensities, with a slightly longer rise time at the output, arising from the delayed triggering of the pump antenna by the incoming spin wave rather than direct microwave drive. This delay can be actively tuned by adjusting the pump power.

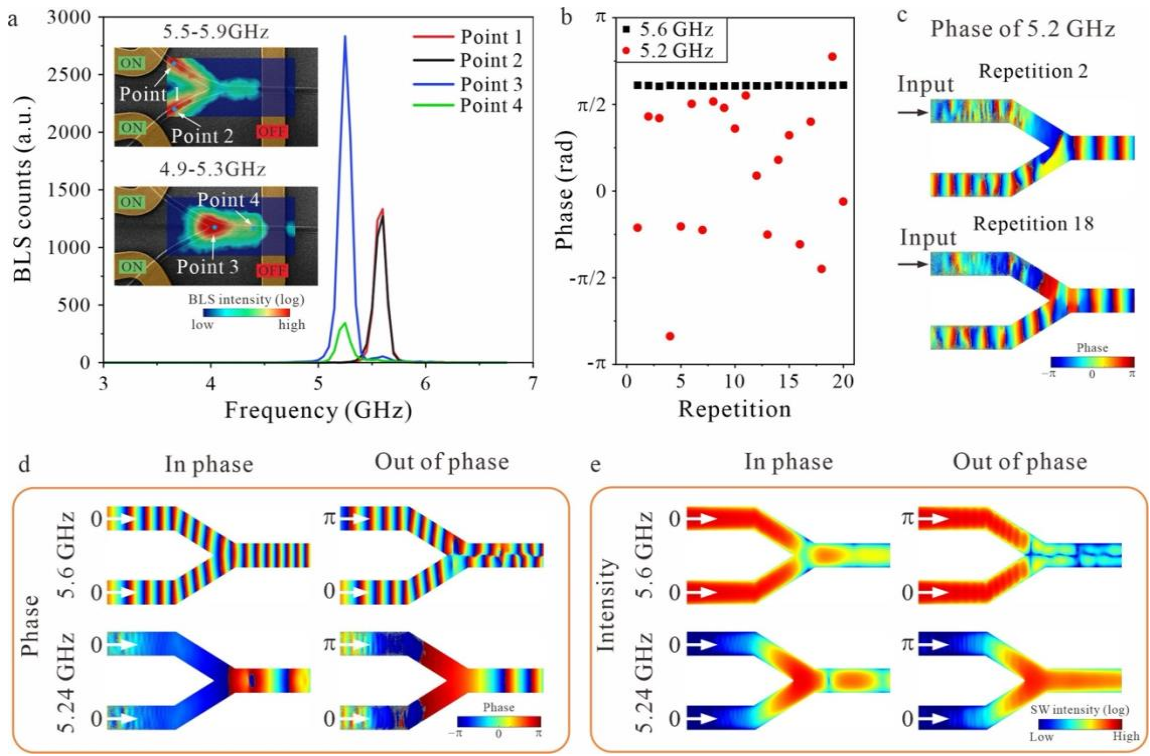

*Figure 4 Nonlinear phase adjustment in a Y-shaped magnonic logic gate.* a, BLS spectra measured at different positions within the gate structure. Insets show 2D maps of the integrated BLS intensity within the indicated frequency ranges, recorded under simultaneous excitation of both inputs. b, Micromagnetic simulations of the spin-wave phase at 5.6 GHz (black dots) and 5.2 GHz (red dots), extracted from regions near the input and the combining area,

*respectively, for single-input excitation repeated twenty times. c, Simulated 2D phase distributions at 5.2 GHz with only the upper input excited for different repetitions, revealing the stochastic nature of the nonlinear phase evolution. d, Simulated 2D phase distributions at 5.6 GHz (top) and 5.24 GHz (bottom) for in-phase and out-of-phase input conditions. e, Corresponding simulated 2D spin-wave intensity maps for the same conditions as in d, showing that while the 5.6 GHz component depends on the input phase, the 5.24 GHz signal remains phase-insensitive.*

As shown in Fig. 3a, the gate produces a stable output that is insensitive to the relative phase of the inputs. To uncover the physical origin of this robustness, we measured the BLS spectra at the four positions indicated in Fig. 4a while simultaneously driving both inputs at 5.6 GHz. Near the input antennas (points 1 and 2), the spectra display the expected peak at the driving frequency (red and black curves). In striking contrast, the spectra in the combining region (point 3) and the output waveguide (point 4) are dominated by a lower-frequency component at approximately 5.25 GHz (blue and green curves). The emergence of this lower-frequency population is attributed primarily to four-magnon scattering of the input magnons, although additional nonlinear effects, such as self-phase modulation at the leading edge of the wave packet, may also contribute [39]. The strong nonlinear scattering and pronounced frequency downshift are enabled by the geometrical nonuniformity of the structure, which provides additional momentum components through its spatial spectrum and thereby relaxes momentum-conservation constraints, allowing multiple scattering channels to occur. By contrast, in a straight, uniform waveguide, four-magnon scattering predominantly generates spectrally nearby modes dictated by the dispersion relation, typically resulting only in a weak spectral broadening of the signal magnon peak (modulation instability) [40]. Furthermore, the wider waveguide supports multiple propagating modes, enabling intermodal scattering that could significantly impact the nonlinear effects [41]. Consistent with this qualitative interpretation, two-dimensional BLS intensity maps show that the high-frequency, initially excited component (5.5-5.9 GHz) dominates the region between the input antennas up to the Y-connection, whereas a pronounced low-frequency component (4.9-5.3 GHz) emerges only after the spin waves enter the wider combining region.

A crucial consequence of this nonlinear frequency conversion is the emergence of a stochastic phase in the low-frequency component that is uncorrelated with the phase of the input signal. Figure 4b shows the simulated phase of magnons at 5.6 GHz (black) and at 5.2 GHz (red), obtained from twenty independent micromagnetic simulations with only the upper input activated and thermal magnons included. The phase of the 5.6 GHz component, evaluated near the input antenna, is identical across all runs because it is imposed by the coherent microwave drive. By contrast, the phase of the 5.2 GHz component, sampled in the combining region, varies randomly from run to run, reflecting the stochastic nature of nonlinear magnon-magnon scattering [42,43].

This behaviour is further illustrated in Fig. 4c, which presents two representative 2D phase maps at 5.2 GHz exhibiting entirely different spatial phase textures. Because only the upper input is excited, the nonlinearly excited 5.2 GHz signal in the upper input arm appears largely noise-like, indicating that nonlinear scattering has not yet developed in this region. In contrast, a well-defined 5.2 GHz phase appears in the output waveguide due to magnon scattering in the combining region, whereas the corresponding signal in another input arm results from reflection of the scattered magnons. The corresponding simulated 2D intensity maps, provided in Supplementary Materials S3, further support this interpretation.

When both inputs are activated, this initially stochastic phase of the low-frequency scattered magnons plays a central role in producing a phase-independent low-frequency output. Once spin waves from one input enter the Y-connection and scatter into the lower frequency state with a random initial phase, subsequent scattering events, including scattering of waves arriving from another input, preferentially lock to this phase via stimulated scattering [see Supplementary Materials S4]. As a result, even when the input waves arrive out of phase, the newly generated low-frequency magnons synchronize to each other and therefore interfere constructively, independent of the relative input phase. This mechanism is illustrated in Fig. 4d: for the 5.6 GHz component (top), out-of-phase excitation produces a nonuniform phase distribution across the waveguide width, indicating mode conversion. In contrast, the 5.24 GHz component (bottom) shows constructive interference for both in-phase and out-of-phase excitation. The corresponding intensity maps in Fig. 4e reveal the same trend: the 5.6 GHz intensity depends strongly on the relative input phase, whereas the 5.24 GHz signal remains nearly identical in both cases.

Consequently, the pump antenna is predominantly activated by the phase-insensitive 5.24 GHz magnons, producing an output that is intrinsically immune to the input phase variations. Experimentally, the low-frequency component is observed to carry more intensity than the original 5.6 GHz excitation at point 4 of the output waveguide (Fig. 4a), highlighting the efficiency of the nonlinear scattering process within Y-connection, where translational symmetry is broken. Crucially, the switching is governed by a threshold-based nonlinear response: although the combined input intensity may be weakly modulated by the relative input phase, the pump antenna is triggered as long as the local intensity exceeds the activation threshold. Once switched on, it emits a fixed, saturated output, thereby eliminating any residual phase-dependent variations.

Additional simulations further show that the similar nonlinear, phase-independent and intensity-normalized operation persists when the device dimensions are scaled down to waveguides only 30 nm in width, demonstrating that the proposed logic architecture remains effective at substantially reduced lateral dimensions [see Supplementary Materials S5]. Together, these results show that the two essential prerequisites for cascadability of intensity-coded operations - intensity normalization and robustness against fabrication induced phase error - are intrinsically fulfilled through nonlinear spin-wave interactions in the combining region.

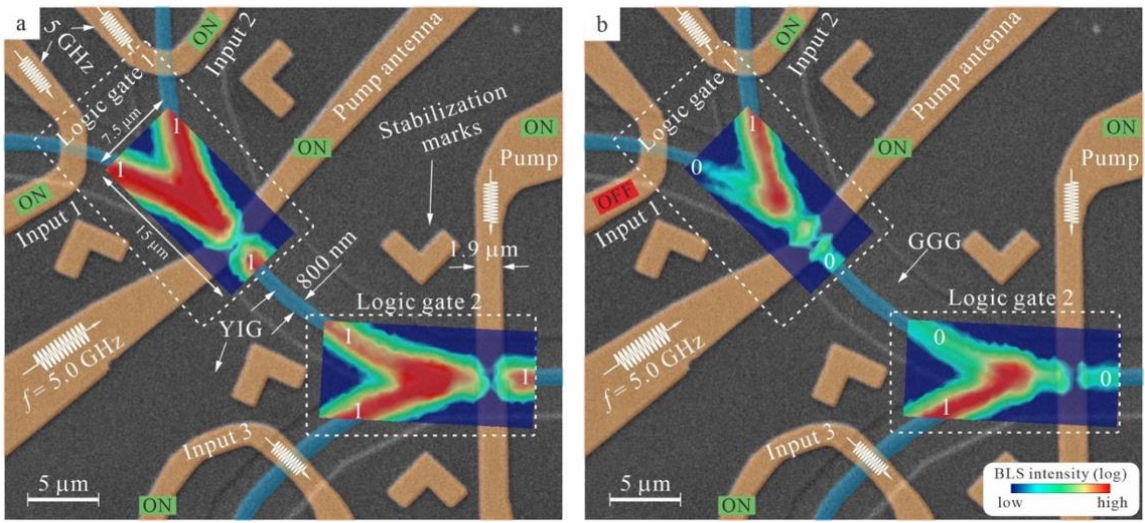

*Figure 5.* ***Two cascaded AND magnonic gates.*** *a, SEM image of two cascaded Y-shaped magnonic logic gates (highlighted in blue), where the output of logic gate 1 directly connected to logic gate 2 as one of its inputs. Five stripe antennas are patterned in gold: three serve as inputs, and the remaining two act as pump antennas. An external magnetic field of 330 mT is applied along the out-of-plane (z-axis) direction. The microwave frequency is fixed at 5.0 GHz, determined by the reduced thickness of the YIG film. The width of the YIG waveguide and antennas are 800 nm and 1.9 μm, respectively. The colormaps display the two-dimensional BLS intensity, integrated over the frequency*

*range 4.6 - 5.5 GHz, separately scanned for both logic gates under the condition that all three input antennas are active. b, Same as in a, but with only input 2 and input 3 activated.*

Based on the physical mechanism discussed above, we present a cascaded magnonic circuit comprising two Y-shaped logic gates patterned from a 47-nm-thick YIG film [28,29] using electron-beam lithography and etching [see Methods]. Unlike in the previous samples, only the YIG regions adjacent to the waveguides were removed, while the rest of the film remained intact. Five stripe antennas were deposited on top of the YIG: three serving as inputs and two as pump antennas (Fig. 5a). The dashed outline highlights the two Y-shaped gates, with the output of the first gate directly feeding into the second as one of its inputs. The operating frequency was reduced to 5.0 GHz due to the increased demagnetizing field in the thinner YIG structure. Input and pump antenna powers were fixed at 5 dBm and 1 dBm, respectively, ensuring that the input antennas directly excite spin waves, while the pump antennas are activated only when spin waves from both inputs arrive simultaneously.

Due to the limited scanning area, BLS measurements were carried out over $15 \times 7.5$ μm² regions, each covering the core of one gate. Figure 5a shows the configuration where inputs 1 and 2 of the first gate are active, while input 3 of the second gate is also switched on. The 2D BLS intensity, integrated from 4.6 to 5.5 GHz, reveals a pronounced spin-wave intensity at the output of the first gate, corresponding to logic '1'. This spin wave propagates through the curved waveguide and enters the second gate as one of its inputs. Simultaneously, input 3 of the second gate is active. Under the combined excitation from gate 1 and input 3, the second pump antenna is triggered, producing a high-intensity output from the second gate, again corresponding to logic '1'. Figure 5b presents the complementary case in which input 1 of the first gate is switched off. The resulting output of the first gate, serving as one input to the second, exhibits only a weak spin-wave intensity. As a consequence, the excitation reaching the second pump antenna remains below threshold, and the second gate yields a low-intensity output corresponding to logic '0'. Together, these measurements demonstrate a fully cascaded magnonic logic operation. They establish a clear pathway for integrating additional gates into larger, more complex magnonic architectures, marking an important advance toward scalable wave-based information processing and on-chip magnonic computing.

**Discussion**

We realised an integrated magnonic chip in nanoscale YIG that delivers phase-insensitive logic with self-normalised outputs and reliable gate-to-gate cascading. AND, OR and three-input majority gates were implemented, and deterministic state transfer is demonstrated in a two-cascade circuit, directly addressing the long-standing challenge of phase sensitivity and signal restoration at the device level. A natural path forward is complementarity with CMOS. A chiplet approach - in which a YIG magnonic die could be co-packaged with a silicon-based controller and memory - would enable short electrical links to on-package transducers while retaining timing, memory and power management in mature CMOS technology. Key enablers are low-loss charge-magnon transducers [34, 44-46], bonding of high-quality YIG films to silicon interposers without degrading damping, and simple cell-level pinouts that allow magnonic gates to integrate into mixed-signal design flows.

Beyond Boolean logic, AI-driven inverse design [13-17] offers a powerful route to compress complex wave-based operations into compact functional elements. By tailoring dispersion, nonlinearity and coupling, magnonic devices could perform tasks such as pattern or sound recognition in a single physical step, with nanosecond-scale latency. Combined with the phase-robust, cascaded logic demonstrated here, this positions magnonic blocks as functional accelerators within hybrid silicon systems. Taken together, these results mark a transition - from magnonics as an academically compelling concept to a technology platform with realistic prospects for large-scale, integrated magnonic chips.

**Method**

**YIG Nanoscale structures fabrication**

The 103-nm- and 47-nm-thick YIG films were grown on 500-μm-thick (111)-oriented gadolinium gallium garnet (GGG) substrates by liquid-phase epitaxy (LPE) [28,29]. The material parameters of the unpatterned films were characterized using stripline vector network analyser ferromagnetic resonance and BLS spectroscopy, yielding a saturation magnetization of $M_s = (140.7 \pm 2.8)$ kA/m, a Gilbert damping parameter of $\alpha = (2 \pm 0.08) \times 10^{-4}$, an inhomogeneous linewidth broadening of $\mu_0 \Delta H_0 = (0.18 \pm 0.01)$ mT, and an exchange constant of $A_{\text{ex}} = (4.22 \pm 0.21)$ pJ/m. These values are representative of high-quality thin-film YIG [28,29,47].

The individual Y-shaped YIG logic gates (Fig. 1) were fabricated from a 103-nm-thick YIG film using a Cr/Ti hard mask and ion-beam milling, following the procedure described in Ref. [47]. The majority gate and the cascaded Y-shaped logic gates were fabricated from a 47-nm-thick YIG film, where a positive electron-beam resist (AR-P 6200.09) was used directly as the etching mask during ion-beam milling. Because a positive-tone resist process was employed, only the device perimeter needed to be exposed, significantly reducing electron-beam lithography time. As a result, during the subsequent ion-beam etching, the YIG was completely removed only in the vicinity of the patterned devices, while the surrounding regions remained covered by the unetched YIG film. The stripe antennas with widths of 2-3 μm, were patterned using standard electron-beam lithography. It consists of ~ 100 nm thick gold and 10 nm thick titanium layer (for adhesion).

**BLS measurements**

A single-frequency laser with a wavelength of 457 nm was focused onto the sample using a 100× microscope objective with a numerical aperture of NA = 0.8. The laser power at the sample was approximately 3 mW. A uniform out-of-plane magnetic field of 330 mT was applied using a 70-mm-diameter NdFeB permanent magnet. Microwave signals with various powers and frequencies were supplied to the antennas to directly excite or amplify spin waves.

For two-dimensional BLS mapping, the sample was scanned relative to the fixed laser spot in steps of a few hundred nanometres using a piezoelectric positioning stage. Stabilization marks fabricated on the sample surface ensured precise realignment of the laser spot, allowing consistent positioning throughout the long measurement cycles.

**Micromagnetic simulations**

The micromagnetic simulations were performed using the GPU-accelerated package MuMax3, incorporating both exchange and dipolar interactions to compute the full space- and time-dependent magnetization dynamics of the investigated structures [48]. Material parameters corresponding to nanometre-thick YIG films were used [28,29,47]: a saturation magnetization of $M_s = 1.407 \times 10^5$ A/m and an exchange constant of $A = 4.2$ pJ/m. The Gilbert damping parameter was set to $\alpha = 7 \times 10^{-4}$ to effectively account for the inhomogeneous linewidth, which cannot be included directly in MuMax3. To suppress spin-wave reflections at the device boundaries, the damping was exponentially increased to 0.5 near the ends of the structure.

The simulation mesh size was defined as $20 \times 20 \times 103$ nm$^3$ (with a single cell along the thickness) for the YIG waveguide. An external magnetic field of $B_{\text{ext}} = 330$ mT was applied along the out-of-plane direction, providing full saturation of the film. To reproduce the stochastic phase behaviour of the scattered magnons, thermal fluctuations were included, and a randomized thermal seed (using thermseed() in MuMax3) was applied for each simulation run [49].

**Data availability**

The data that support the plots presented in this paper are available from the corresponding authors upon reasonable request.

**Code availability**

The code used to analyse the data and the related simulation files are available from the corresponding author upon reasonable request.

**References**

1. B. Flebus, D. Grundler, B. Rana, Y. Otani, I. Barsukov, et al., The 2024 magnonics roadmap, *J. Phys: Condens. Matter* **36**, 363501 (2024).

2. Q. Wang, G. Csaba, R. Verba, A. V. Chumak, and P. Pirro, Nanoscale magnonic networks, *Phys. Rev. Applied* **21**, 040503 (2024).

3. A. V. Chumak, V. I. Vasyuchka, A. A. Serga, B. Hillebrands, Magnon spintronics. *Nat. Phys.* **11**, 453-461 (2015).

4. B. Dieny, I. L. Prejbeanu, K. Garello, P. Gambardella, P. Freitas et al., Opportunities and challenges for spintronics in the microelectronic industry (Topical Review). *Nat. Electron.* **3**, 446-459 (2020).

5. A. V. Chumak, P. Kabos, M. Wu, C. Abert, C. Adelmann, *et al*., Advances in Magnetics Roadmap on Spin-Wave Computing, IEEE Trans. Magn. **58**, 1 (2022).

6. Y. Kajiwara, K. Harii, S. Takahashi, J. Ohe, K. Uchida, et al., Transmission of electrical signal by spin-wave interconversion in a magnetic insulator. *Nature* **464**, 262 (2010).

7. J. Han, P. Zhang, J. T. Hou, S. A. Siddiqui, L. Liu, Mutual control of coherent spin waves and magnetic domain walls in a magnonic device, *Science* **336**, 1121 (2019).

8. Q. Wang, M. Kewenig, M. Schneider, R. Verba, F. Kohl, et al., A magnonic directional coupler for integrated magnonic half-adders. *Nat. Electron.* **3**, 765 (2020).

9. A. Kumar, A. K. Chaurasiya, V. H. González, N. Behera, A. Alemán, et al., Spin-wave-mediated mutual synchronization and phase tuning in spin Hall nano-oscillators, *Nat. Phys.* **21**, 245 (2025).

10. L. Sheng, A. Duvakina, H. Wang, K. Yamamoto, R. Yuan, et al., Control of spin currents by magnon interference in a canted antiferromagnet, *Nat. Phys.* **21**, 740 (2025).

11. H. Qin, R. B. Holländer, L. Flajšman, F. Hermann, R. Dreyer, et al., Nanoscale magnonic Fabry-Pérot resonator for low-loss spin-wave manipulation, *Nat. Commun.* **12**, 2293 (2021).

12. S. Wintz, V. Tiberkevich, M. Weigand, J. Raabe, J. Lindner, et al. Magnetic vortex cores as tunable spin-wave emitters. *Nat. Nano.*, **11**, 948-953 (2016).

13. Q. Wang, A. Chumak, & P. Pirro, Inverse-design magnonic devices, Nat. Commun. 12, 2636 (2021).

14. N. Zenbaa, F. Majcen, C. Abert, F. Bruckner, N. J. Mauser, et al., Realization of inverse-design magnonic logic gates, *Sci. Adv.* **11**, eadu9032 (2025).

15. Á. Papp, W. Porod, & G. Csaba, Nanoscale neural network using non-linear spin-wave interference, *Nat. Commun.* **12**, 6422 (2021).

16. L. Körber, C. Heins, T. Hula, J.-V. Kim, S. Thlang, H. Schultheiss, J. Fassbender, & K. Schultheiss, Pattern recognition in reciprocal space with a magnon-scattering reservoir, *Nat. Commun.* **14**, 3954 (2023).

17. D. Breitbach, M. Bechberger, H. Mortada, B. Heinz, R. Verba, et al., All-magnonic neurons for analog artificial neural networks, arXiv:2509.18321 (2025).

18. K. -S. Lee, S. -K. Kim Conceptual design of spin wave logic gates based on a Mach–Zehnder-type spin wave interferometer for universal logic functions, *J. Appl. Phys.* **104**, 053909 (2008)

19. M. Balynsky, D. Gutierrez, H. Chiang, A. Kozhevnikov, D. Dudko et al., A Magnetometer Based on a Spin Wave Interferometer, *Sci. Rep.* **7**, 11539 (2017).

20. T. Schneider, A. A. Serga, B. Leven, B. Hillebrands, R. L. Stamps, M. P. Kostylev, Realization of spin-wave logic gates, *Appl. Phys. Lett.* **92**, 022505 (2008).

21. A. V. Chumak, A. A. Serga, B. Hillebrands, Magnon transistor for all-magnon data processing, *Nat. Commun.* **5**, 4700 (2014).

22. G. Talmelli, T. Devolder, N, Träger, J. Förster, S. Wintz, et al., Reconfigurable submicrometer spin-wave majority gate with electrical transducers, *Sci. Adv.* **6**, eabb4042 (2020).

23. H. Merbouche, B. Divinskiy, D. Gouéré, R. Lebrun, A. E. Kanj, et al. True amplification of spin waves in magnonic nano-waveguides, *Nat. Commun.* **15**, 1560 (2024).

24. T. Fischer, M. Kewenig, D. A. Bozhko, A. A. Serga, I. I. Syvorotka, et al., Experimental prototype of a spin-wave majority gate, *Appl. Phys. Lett.* **110**, 152401 (2017).

25. R. Verba, M. Carpentieri, G. Finocchio, V. Tiberkevich, & A. Slavin, Amplification and stabilization of large-amplitude propagating spin waves by parametric pumping. *Appl. Phys. Lett*. **112**, 042402 (2018).

26. T. Brächer, F. Heussner, P. Pirro, T. Fischer, M. Geilen, et al., Time- and power-dependent operation of a parametric spin-wave amplifier. *Appl. Phys. Lett.* **105**, 232409 (2014).

27. Q. Wang, R. Verba, K. Davídková, B. Heinz, S. Tian, et al., All-magnonic repeater based on bistability. *Nat. Commun.* **15**, 7577 (2024).

28. C. Dubs, O. Surzhenko, R. Linke, A. Danllewsky, U. Brückner and J. Dellith, Sub-micrometer yttrium iron garnet LPE films with low ferromagnetic resonance losses, J. Phys. D: Appl. Phys. 50, 204005 (2017).

29. C. Dubs, O. Surzhenko, R. Thomas, J. Osten, T. Schneider, et al., Low damping and microstructural perfection of sub-40nm-thin yttrium iron garnet films grown by liquid phase epitaxy. *Phys. Rev. Materials* **4**, 024426 (2020).

30. Y. Li, V. V. Naletov, O. Klein, J. L. Prieto, M. Muñoz et al., Nutation Spectroscopy of a Nanomagnet Driven into Deeply Nonlinear Ferromagnetic Resonance. *Phys. Rev. X* **9**, 041036 (2019).

31. Q. Wang, R. Verba, B. Heinz, M. Schneider, O. Wojewoda, et al. Deeply nonlinear excitation of self-normalized short spin waves, *Sci. Adv.* **9**, eadg4609 (2023).

32. Y. Wang, G. Zhang, D. Zhang, T. Li, C.-M Hu, et al., Bistability of Cavity Magnon Polaritons, *Phys. Rev. Lett.* **120**, 057202 (2018).

33. R. Shen, J. Li, Z. Fan, Y. Wang, and J. Q. You, Mechanical Bistability in Kerr-modified Cavity Magnomechanics, *Phys. Rev. Lett.* **129**, 123601 (2022).

34. F. Vanderveken, V. Tyberkevych, G. Talmelli, B. Sorée1, F. Ciubotaru, C. Adelmann, Lumped circuit model for inductive antenna spin-wave transducers, *Sci. Rep.* **12**, 3796 (2022).

35. M. Sushruth, M. Grassi, K. Ait-Oukaci, D. Stoeffler, Y. Henry et al., Electrical spectroscopy of forward volume spin waves in perpendicularly magnetized materials, *Phys. Rev. Res.* **2**, 043203 (2020).

36. T. Sebastian, K. Schultheiss, B. Obry, B. Hillebrands, and H. Schultheiss, Micro-focused Brillouin light scattering: imaging spin waves at the nanoscale. *Front. Phys.* **3**, 55 (2015).

37. F. Ciubotaru, G. Talmelli, T. Devolder, O. Zografos, M, Heyns, C. Adelmann, First experimental demonstration of a scalable linear majority gate based on spin waves, in Proceedings of the 2018 IEEE International Electron Device Meeting (IEDM), 36.1 (2018)

38. A. C. Bunea, F. Ciubotaru, F. Meng, C. Adelmann, O. G. Profirescu, and D. Neculoiu, Investigation of electromagnetic feedthrough in spin wave based majority gates, International Semiconductor Conference (CAS), Sinaia, Romania, 2024, pp. 101–104.

39. M. Merbouche, B. Divinskiy, K. O. Nikolaev, C. Kaspar, W. H. Pernice, et al., Giant nonlinear self-phase modulation of large-amplitude spin waves in microscopic YIG waveguides, *Sci. Rep.* **12**, 7246 (2022).

40. S. Jun, S. A. Nikitov, R. Marcelli, and R. De Gasperis, Parametric and modulation instabilities of magnetostatic surface spin waves in ferromagnetic films, *J. Appl. Phys.* **81**, 1341 (1997).

41. M. Mohseni, Q. Wang, B. Heinz, M. Kewenig, M. Schneider, et al., Controlling the Nonlinear Relaxation of Quantized Propagating Magnons in Nanodevices, *Phys. Rev. Lett.* **126**, 097202 (2021).

42. V. S. L'vov, Wave Turbulence Under Parametric Excitation, (Springer, 1994).

43. M. Koster, M. R. Schweizer, T. Noack, V. I. Vasyuchka, D. A. Bozhko, et al., Spontaneous Emergence of Phase Coherence in a quasiparticle Bose-Einstein Condensate, arXiv:2507.16862 (2025).

44. R. Erdélyi, G. Csaba, L. Maucha, F. Kohl, B. Heinz et al., Design rules for low-insertion-loss magnonic transducers, *Sci. Rep.* **15**, 9806 (2025).

45. F. Bruckner, K. Davídková, C. Abert, A. Chumak & D. Suess, Micromagnetic simulation and optimization of spin-wave transducers, *Sci. Rep.* **15**, 19993 (2025).

46. F. Kohl, B. Heinz, M. Wagner, C. Adelmann, F. Ciubotaru, & P. Pirro, Modelling spin-wave interference with electromagnetic leakage in micron-scaled spin-wave transducers, arXiv:2509.24647 (2025).

47. B. Heinz, T. Brächer, M. Schneider, Q. Wang, B. Lägel, et al. Propagation of spin-wave packets in individual nanosized yttrium iron garnet magnonic conduits. *Nano Lett.* **12**, 4220 (2020).

48. A. Vansteenkiste, J. Leliaert, M. Dvornik, M. Helsen, F. Garcia-Sanchez, et al. The design and verification of MuMax3. *AIP Adv.* **4**, 107133 (2014).

49. J. Leliaert, J. Mulkers, J. De Clercq, A. Coene, M. Dvornik, & B. Van Waeyenberge, Adaptively time stepping the stochastic Landau-Lifshitz-Gilbert equation at nonzero temperature: Implementation and validation in MuMax3, *AIP Adv*, **7**, 125010 (2017).

**Acknowledgements**

This work was supported from the National Natural Science Foundation of China (Grant No. 12574118). R. V. acknowledge support by the NAS of Ukraine, project #0124U000270. P. P. acknowledges support by the Deutsche Forschungsgemeinschaft (DFG, German Research Foundation) -TRR 173–268565370 ("Spin + X", Project B01) and by the European Research Council within the Starting Grant No. 101042439 "CoSpiN". A. V. C. acknowledges the financial support of the Austrian Science Fund (FWF) by means of grant MagNeuro no. 10.55776/PIN1434524. C. D. acknowledges the financial support by the Deutsche Forschungsgemeinschaft (DFG, German Research Foundation) - 271741898.

**Author Contributions**

M. G. performed BLS measurements and micromagnetic simulations with help from X. G., Z. Z.. X. J. and K. D. fabricate the nanoscale sample with help from Y. R., K. C. and J. L.. R. V. provided theoretical support and analysis. C. D. grew the YIG films. A. C., P. P., and Q. W. led this project. Q. W. conceived the idea and wrote the manuscript with the help of all the coauthors. All authors contributed to the scientific discussion and commented on the manuscript.

**Competing Interests**

The authors declare no competing interests.

**Correspondence** and requests for materials should be addressed to Q. W.